\documentclass[twocolumn,showpacs,prb,groupedaddress]{revtex4}

\usepackage{graphicx}
\usepackage{dcolumn}
\usepackage{bm}

\begin{document}

\preprint{PRB - BJR740B - resub}

\title{Planar $^{17}$O NMR study of Pr$_y$Y$_{1-y}$Ba$_2$Cu$_3$O$_{6+x}$}

\author{W.A. MacFarlane}
 \altaffiliation[Current Address~:~]{Department of Chemistry, University of British Columbia, Vancouver, Canada V6T 1Z1.}
\author{J. Bobroff}
\author{P. Mendels}
\author{L. Cyrot}
\author{H. Alloul}
\author{N. Blanchard}
\affiliation{%
LPS, B\^{a}t. 510, Universit\'e Paris-Sud, UMR8502 CNRS, 91405, Orsay Cedex, France
}%
\author{G. Collin}
\affiliation{
LLB, CEN Saclay, CEA-CNRS, 91191 Gif-sur-Yvette, France
}%
\author{J.-F. Marucco}
\affiliation{
LCNS, Univerist\'e Paris-Sud, 91405, Orsay, France
}%

\date{\today}

\begin{abstract}
We report the planar $^{17}$O NMR shift in Pr substituted
YBa$_{2}$Cu$_{3}$O$_{6+x}$, which at $x=1$ exhibits a characteristic pseudogap
temperature dependence, confirming
that Pr reduces the concentration of mobile holes in the CuO$_{2}$ planes.
Our estimate of the rate of this counterdoping effect,
obtained by comparison with the shift in pure samples with reduced oxygen content,
is found insufficient to explain the observed reduction of T$_c$.
From the temperature dependent magnetic broadening of the $^{17}$O NMR we
conclude that the Pr moment and the local magnetic defect induced in the
CuO$_2$ planes produce a long range spin polarization in the planes,
which is likely associated with the extra reduction of T$_c$.
We find a qualitatively different behaviour in the oxygen depleted
Pr$_y$Y$_{1-y}$Ba$_2$Cu$_3$O$_{6.6}$, i.e. the suppression of T$_c$ is nearly the same,
but the magnetic broadening of the $^{17}$O NMR appears weaker.
This difference may signal a weaker coupling of the Pr to the planes
in the underdoped compound, which might be linked with the larger Pr
to CuO$_2$ plane distance, and correspondingly weaker hybridization.
\end{abstract}

\maketitle

\typeout{74.62.Dh Substitutions, impurities} \typeout{76.60.-k NMR}
\typeout{74.25.Ha Superconductivity and Magnetism}

Among the isostructural Rare Earth analogues of the well-studied
high-T$_{c}$ system, YBa$_{2}$Cu$_{3}$O$_{6+x}$, only Pr/Y substitution
suppresses superconductivity. The most widely accepted explanation of this
effect is that it is due to counterdoping of the planes, i.e. the
concentration of mobile holes in the planes is depleted, resulting in the
{\it insulating} behaviour found for $y\stackrel{>}{\sim }0.6$ (e.g. Ref.~
\cite{maprev}). Several proposals for the origin of this counterdoping have
been advanced; for example, the valence of Pr may be larger than 3$+$ (e.g.
Ref. \cite{motoi}), the large ionic radius of Pr$^{3+}$ may influence the
chain-plane charge transfer (similar to other large trivalent ions\cite{lutg}
), or strong hybridization of Pr with its neighbouring planar oxygen may
cause holes to be removed from the conduction band into localized\cite{FR}
or insulating-band\cite{LM} states. X-Ray and Cu NQR experiments\cite
{merzohno} apparently rule out the first two possibilities.

Planar $^{17}$O NMR provides a sensitive probe of the spin susceptibility ($
\chi _{s}$) of the CuO$_{2}$ planes in the cuprate high T$_{c}$
superconductors. In particular, in YBa$_{2}$Cu$_{3}$O$_{6+x}$, $\chi _{s}(T)$
has been shown to be strongly dependent on the carrier content ($n$) of the
planes, especially near T$_{c}$-optimal doping\cite{miami}.
In the
present work, we argue that determining $\chi _{s}(T)$
provides a measure of $n$ in the normal state of
Pr$_{y}$Y$_{1-y}$Ba$_{2}$Cu$_{3}$O$_{7}$ (Pr$_{y}$:O$_{7}$).
Such a local measurement of $n,$ which is free of any
influence of parasitic phases, quantitatively constrains models of
counterdoping. Furthermore, we find a Curie-like line broadening of
the $^{17}$O NMR indicating that the Pr ion is a magnetic
perturbation which induces a long range oscillatory spin polarization
response in the CuO$_{2}$ planes, as has been found for Ni and Zn
in-plane substitutions\cite{plimp}. This effect is somewhat analogous to the RKKY
oscillations found in conventional metals containing magnetic impurities.
Such a perturbation may be responsible for the extra reduction of T$_{c},$
which is obtained after removing the net counterdoping effect.

Finally, motivated by results from underdoped YBa$_{2}$Cu$_{3}$O$_{6+x}$
with Zn and Ni as in-plane Cu substitutions, which
exhibit enhanced local magnetic effects and T$_{c}$ reduction
relative to optimal doping\cite{plimp}, we have also studied deoxygenated
samples (Pr$_{y}$:O$_{6.6}$).
The results indicate that Pr
induces a weaker magnetic perturbation and is therefore less coupled to the
CuO$_{2}$ planes in Pr$_{y}$:O$_{6.6}$.

After detailing the samples and technical aspects of the measurement, we
present the results for Pr$_{y}$:O$_{7}$ then Pr$_{y}$:O$_{6.6}$. Then we
discuss the data in relation to the results of other experiments.

\section{Samples}

The powder samples were synthesized by conventional high temperature solid
state reaction of Pr$_{6}$O$_{11}$, Y$_{2}$O$_{3}$, BaO$_{2}$ and CuO as
detailed in Ref. \cite{grevin}.
This synthesis procedure reproducibly results in a family
Pr$_{y}$Y$_{1-y}$Ba$_{2}$Cu$_{3}$O$_{7},$ of compounds with
monotonically decreasing $T_{c}(y)$\cite{refzou}. Pr is not expected
to affect the maximal oxygen uptake of the sample significantly\cite{horn},
and this was confirmed thermogravimetrically, i.e. $\delta x\leq 0.008$ for $%
y=0.2$ relative to $y=0$. The presence of Pr substituted on the Ba site can
be detected by Nuclear Quadrupolar Resonance (NQR) of the chain Cu as has
been established by intentional Pr/Ba substitution\cite{grevin2}. The chain
NQR spectrum for samples synthetized by us with the same procedure, but with
much larger Pr content is the same as that of pure YBa$_{2}$Cu$_{3}$O$_{7}$,
indicating that for our samples Pr/Ba substitution is {\it negligible}. The
samples were isotopically enriched by annealing in a $^{17}$O$_{2}$ enriched
atmosphere. Subsequent deoxidization was accomplished by vacuum annealing
under thermogravimetric control. The samples were aligned by suspension in
Stycast 1266 epoxy cured in a 7.5 T magnetic field. Alignment was checked
using $^{89}$Y NMR at 300 K. It was found that for $y\leq 0.1$, the samples
were well aligned, but for $y=0.2$, the powder did not completely align.
However, systematic differences in the results due to imperfect alignment
are negligible in comparison to the $n$ dependent contribution of $\chi
_{s}(T)$ \cite{why}.

\section{NMR Technical Aspects}

\subsection{Experimental Method}

Conventional spin echo NMR spectra of the central transition of the plane
oxygen (O(2,3)) were taken in the temperature range 80 to 350 K with the
field aligned parallel to the c-axis (except as noted above). The typical $%
\pi /2$-$\tau $-$\pi $ sequence consisted of $\sim 1\mu $s $\pi /2$ pulse
and a $\tau $ of $\sim 100\mu \text{s}\ll T_{2}$. The contrast in spin
lattice relaxation rates ($T_{1}^{-1}$) between the plane and apical O(4)
oxygen sites was exploited (using fast repetition $\sim 50$ ms) to eliminate
contamination from O(4). Typical Fourier transformed spectra are illustrated
in Fig.~1. From the raw spectra, the effect of Pr substitution
is clearly to cause temperature dependent shift and broadening of the
resonance. As shown below, the temperature dependent shift is very similar
to deoxidized pure YBCO$_{7}$. We analyze the frequency spectra by
extracting the peak position
and the linewidth $\Delta \nu $ defined as twice the
upper frequency halfwidth at half maximum. This definition of $\Delta\nu $
excludes the contribution to the linewidth of a slight
inhomogeneity of $n$. This inhomogeneity of about $\pm 5 \% $, which is also
found in pure compounds, induces a larger broadening of the low
frequency side of the line.

The average Knight shift $^{17}K_{c}$ was extracted from the peak position
as follows. We correct for the small second order quadrupolar shift of the
central line ($\nu _{1/2}^{(2)}$) assuming the same values of the
quadrupolar interaction as in pure YBa$_{2}$Cu$_{3}$O$_{6+x}$\cite
{yoshi90,taki}. For the field along $c$, $\nu _{1/2}^{(2)}$varies from 12.1
kHz ($x=1$) to 10.6 kHz ($x=0.6$) and is independent of temperature. These
corrections are not large (less than 300 ppm at the experimental field), and
are very weakly doping dependent. Thus, although one does expect changes in
the average quadrupolar interaction with $y$ between the extremes of $y=0$%
\cite{taki} and $y=1$\cite{ko}, such effects are small compared to the $T$
dependent part of the shift.

\begin{figure}[t]
\centerline{\includegraphics[width=7cm]{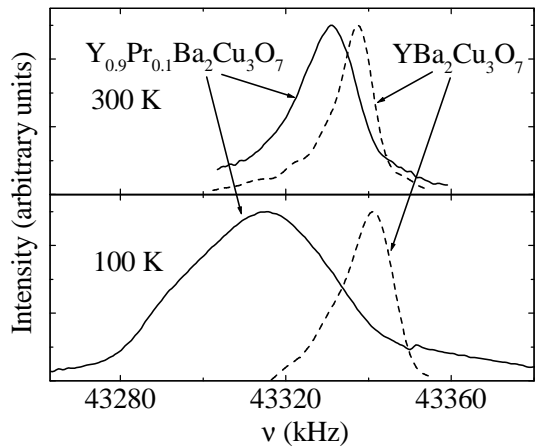}}
\vskip 0.2cm
\caption{ The
$^{17}$O(2,3) NMR mainlines ($I_{z}=-\frac{1}{2}%
\leftrightarrow \frac{1}{2}$ transition) for pure YBCO$_{7}$ (dashed) and Pr$%
_{0.1}$:O$_{7}$ (solid) at 300 K (top panel) and 100 K (bottom panel) from
fast repetition (50 ms) spin echo measurements. The downward shift of the
peak position in the Pr$_{0.1}$:O$_{7}$ case is due to the loss of
susceptibility characteristic of the opening of the pseudogap. The left axis
position corresponds to the zero of shift. The small feature at 43350 kHz is
an artefact of the RF irradiation frequency. The $T$ dependent broadening is
also apparent. }
\label{datafig}
\end{figure}

\subsection{Calibration of $^{17}$K$_c$(2,3)}

In order to compare the shift effect to the doping effect of changing the
oxygen content in pure YBCO$_{6+x}$, a set of curves was developed for $%
0.6\le x\le 1$ to summarize the changes of the shift in terms of the single
parameter $x$, see Fig.~2. The function, based on a well-known
phenomenological pseudogap $T$-dependence\cite{mehring}, was of the form:
\begin{eqnarray}
^{17}K_{c}(T,x)=a(x)-bT+c(1-\tanh {^{2}(T^{*}(x)/T)}),  \label{superspingap}
\end{eqnarray}
where $T^{*}$ is a measure of the pseudogap. The $T$ profile of the shift
varies strongly with $x$ for $x>0.85$ (Fig.~2). The dependence of
the shift is weaker\cite{yoshi90,yoshithes} for $x<0.85$. In particular for $%
0.6<x<0.85$, the main effect is a $T$ independent downward offset of $%
^{17}K_{c}$, while for $0.45<x<0.6$ there is additionally a compression of
the variation of $^{17}K_{c}$, e.g. in the difference between 300 K and 100
K. Thus measurement of the $T$ profile of the shift constitutes a measure of
the doping level of the CuO$_{2}$ planes with high sensitivity near optimal
doping and less sensitivity at lower doping.

\begin{figure}[t]
\centerline{\includegraphics[width=7cm]{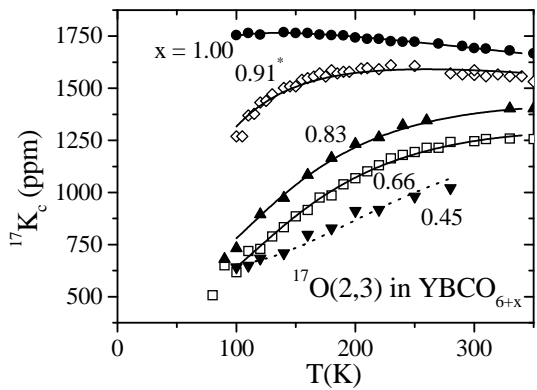}}
\vskip 0.2cm
\caption{ Variation of the plane oxygen temperature dependent Knight shift
for various dopings in YBCO$_{6+x}$. The fit curves are to a function of the
form Eq. (\ref{superspingap}). The $x=0.91$ points are obtained from scaling
the $^{89}$Y shift \protect\cite{takiscale}. The data at $x=0.45$ is from
\protect\cite{yoshithes}. Similar curves have been used to fit the $^{89}Y$
Knight shift \protect\cite{miami}. The shift curve amounts to a very
sensitive measurement of the average doping in the range $x=0.8$ to 1.0.}
\label{calfig}
\end{figure}

\section{Results for {Pr$_y$Y$_{1-y}$Ba$_2$Cu$_3$O$_7$}}

\subsection{Shifts: The Effect on Doping}

For the Pr$_{y}$:O$_{7}$ samples, the measured shifts relative to the $^{17}$%
O frequency in water are shown in Fig.~3 together with the
curves for oxygen depleted pure YBa$_{2}$Cu$_{3}$O$_{6+x}$ (From Fig.~2).
Note that the spectra are featureless, and that we do not resolve
specific sites such as the near neighbours (n.n.) of Pr. However, such
local measurements have been carried out using the $^{89}$Y NMR of the n.n.
sites of Pr\cite{Feri}. These results lead us to anticipate that the $^{17}$%
O n.n. sites of the Pr are wiped out of the present spectra, which are thus
representative of the average $\chi_s$ far from the Pr.
The data from the Pr doped samples can be compared with curves obtained by
interpolation between those of the deoxygenated YBCO samples. The excellent
fits which are obtained indicate that $^{17}K_{c}$ follows the deoxygenated
behaviour quite closely. The fact the Pr data are well-fit by the
calibration curves {\it over the whole temperature range} confirms that
changes in $T$ independent shift contributions due to Pr substitution are
negligible.

In both the
deoxygenated and Pr doped data, the opening of the pseudogap is apparent in
the temperature profile of $^{17}K_{c}$.
While there is no detailed theoretical understanding of the electronic structure
of the doped CuO$_2$ planes (from which one could calculate $\chi _{s}(n,T)$),
the resemblance is so strong (Fig.~3) that it is
quite reasonable to assume a common origin for the evolution of
the $T$ profiles, i.e. a reduction of $n$. Moreover, measurements of quite
different properties (e.g.\cite{bern}) also find evidence of a reduction of $n$.
Henceforth, we will simply assume this is the case and proceed with an analysis
based on the phenomenological model of Eq.~(1).
From the fits, we can thus extract a value for the equivalent oxygen
concentration $x^{\text{eq}}$ as a function of Pr content $y$, which is
reported in Fig.~4 together with a linear fit.

\begin{figure}[t]
\centerline{\includegraphics[width=7cm]{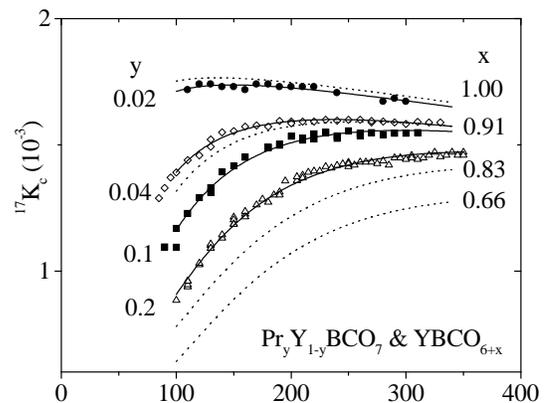}}
\vskip 0.2cm
\caption{
$^{17}$K$_c$ for various Pr concentrations $y$ in fully oxidized
Pr$_y$Y$_{1-y}$Ba$_2$Cu$_3$O$_{7}$ together with (solid) curves for
deoxygenated YBa$_2$Cu$_3$O$_{6+x}$.
The dashed curves are fits to a phenomenological
function indexed by a single parameter
$x^{\text{eq}}$.
}
\label{o7shif}
\end{figure}

\begin{figure}[t]
\centerline{\includegraphics[width=7cm]{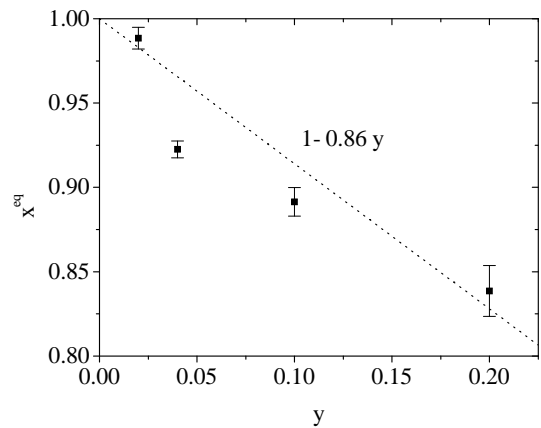}}
\vskip 0.2cm
\caption{ The equivalent oxygenation as a function of Pr concentration $y$
in Pr$_y$Y$_{1-y}$Ba$_2$Cu$_3$O$_{7}$.}
\label{xofy}
\end{figure}

The decrease of T$_{c}$ with $y$ is much more rapid than expected
from counterdoping alone. This can be seen by comparing T$_{c}(x)$
for the oxygen depleted materials with T$_{c}(x^{\text{eq}})$
(Fig.~5).\ The difference between the two curves clearly shows that
another mechanism of T$_{c}$ suppression is present. The excess reduction in
T$_{c}$, $\Delta $T$_{c}^{\text{ex}}$, has been extracted as shown in Fig.~5,
and is plotted as a function of Pr concentration in Fig.~6.
The slope of $\Delta $T$_{c}^{\text{ex}}(y)$ is about 0.51 K/\% .

\begin{figure}[b]
\centerline{\includegraphics[width=7cm]{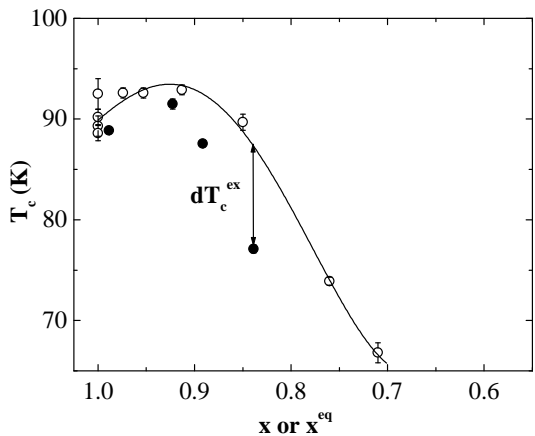}}
\vskip 0.2cm
\caption{ T$_c$ as a function of $x$ for pure YBa$_2$Cu$_3$O$_{6+x}$ (open
points) and of $x^{\text{eq}}$ for Pr$_y$Y$_{1-y}$Ba$_2$Cu$_3$O$_{7}$ (solid
points) where all the samples are from the same source, and all the values
of T$_c$ were obtained by extrapolating the linear mid-transition region of
the field cooled (typically $B=30$ G) DC magnetic susceptibility ($\protect%
\chi$) to the zero of $\protect\chi$. }
\label{tcfig}
\end{figure}

Let us compare this result with former studies.\ Previous Cu NMR
measurements for Pr substituted samples also found underdoped behavior in
both the plane copper shift and spin lattice relaxation temperature
dependence\cite{reyes}. Also using simultaneous Ca/Y and Pr/Y substitution a
somewhat larger slope of $\Delta $T$_{c}^{\text{ex}}(y)$ was inferred\cite
{maprev}. But in both cases there was no direct measurement of the
counterdoping effect of Pr. Here the specific advantages of the $^{17}$O
NMR, such as its high sensitivity and intrinsically narrow resonance,
together with the systematics of our study have allowed us to reach accurate
quantitative conclusions.

\begin{figure}[t]
\centerline{\includegraphics[width=7cm]{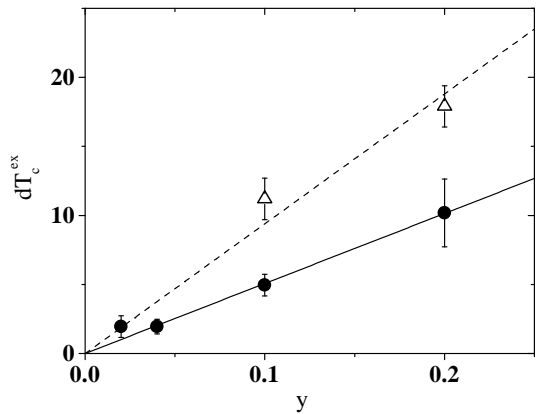}}
\vskip 0.2cm
\caption{ Solid points: The excess reduction of T$_c$ as defined in the
previous figure as a function of Pr concentration $y$ for Pr$_y$Y$_{1-y}$Ba$%
_2$Cu$_3$O$_{7}$. Open points: reduction of T$_c$ from its $y=0$ value
(59.5(1.0) K) in Pr$_y$Y$_{1-y}$Ba$_2$Cu$_3$O$_{6.6}$. }
\label{dtcfig}
\end{figure}

If we extrapolate from the dilute regime of our measurements, the
observed counterdoping effect is already sufficient to explain the
metal-insulator transition at $y\approx 0.6$, where $x^{\text{eq}}$ $\approx
0.4$. But, as the decrease of T$_{c}$ is even larger than that obtained from
the counterdoping alone, we expect a non-superconducting part of the phase
diagram whose ground state is likely an insulating magnetically disordered
state, similar to Zn substituted samples\cite{mendels}.

\subsection{Broadening: Magnetic Effects}

We can gain more insight into the effects induced by Pr substitution and
more specifically on the origin of $\Delta T_{c}^{\text{ex}}$ from the
$^{17} $O {\it linewidths}, $\Delta \nu $. The observed $\Delta \nu $ are
presented in Fig.~7 together with that of the pure system. It is
apparent that there is an excess broadening related to the Pr concentration
which increases as $T\rightarrow 0$. As Pr is a paramagnetic ion with a
local susceptibility possessing a Curie $T^{-1}$ component, it is
natural to associate the line broadening with such a magnetic contribution.
Indeed $\Delta \nu $(T) in Pr$_{y}$:O$_{7}$ scales approximately with the Pr
susceptibility\cite{jaya}. To interpret this broadening, we move now to a
discussion of its origin.

\begin{figure}[t]
\centerline{\includegraphics[width=7cm]{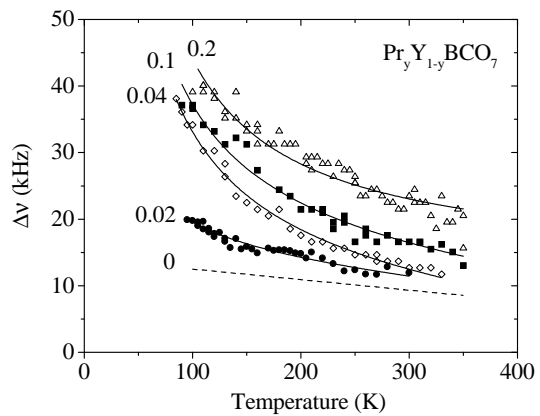}}
\vskip 0.2cm
\caption{ The widths of the O(2,3) mainlines, and the width in pure YBa$_2$Cu%
$_3$O$_{7}$ (dashed). $y=$0.02(circles), 0.04(diamonds), 0.1(squares),
0.2(triangles).}
\label{o7wid}
\end{figure}

The spatial distribution of local fields due to randomly placed magnetic
impurities is expected to give rise to a broadening of the host nuclear
resonance lines which scales with the impurity contribution to the
susceptibility, typically a Curie law for dilute weakly interacting
impurities. In contrast, dilute {\it nonmagnetic} disorder generally
produces a temperature independent broadening due, for example, to a
distribution of electric field gradients.

The local {\em magnetic} field always contains the direct dipolar field of
the impurity moments, but in metallic systems the field distribution sampled
by the nuclei is usually dominated by (RKKY) spin polarization of the
conduction band. Let us first demonstrate that the computed dipolar
fields are negligible compared to the measured broadening. Although
the precise nature of the Pr moment is still controversial\cite{neut}, its
contribution to the macroscopic susceptibility is accounted for (in this T
range) by a $\approx 2.7\mu _{B}$ Curie term plus a constant. In the dilute
limit, the dipolar broadening may be estimated \cite{wal} as
\[
\Delta \nu =\frac{8\pi }{9\sqrt{3}}nD\langle S_{z}\rangle ,
\]
where $n$ is the volume concentration of the randomly placed impurities, $D$
is the dipolar coupling to the nuclei with gyromagnetic ratio $\gamma _{n}$,
i.e. $D=2\mu _{B}\gamma _{n}$, and $\langle S_{z}\rangle $ is the
thermodynamically averaged impurity spin polarization,
proportional to the impurity susceptibility.
Substituting the values for the Pr moment and the oxygen nuclei yields
\[
\Delta \nu =\frac{3840\text{(kHz K)}y}{T}
\]
For 4\% Pr, the variation of this width between 300 K and 80 K is less than
1.4 kHz, more than an order of magnitude smaller than the observed value\cite
{widnot}. Thus the Curie-like broadening of the $^{17}$O\ NMR must be due to
an induced long range spin polarization associated with a coupling of the Pr
magnetic defect to the plane carriers. Such an effect is not observed for
other rare earths for which a dipolar contribution is sufficient to explain
the NMR widths. We discuss in section V the possible origins of this
coupling.

\section{Results for \text{Pr$_y$Y$_{1-y}$Ba$_2$Cu$_3$O$_{6.6}$}}

The reported {\it decrease} in the effectiveness of Pr in reducing T$_{c}$
in Pr$_{y}$:O$_{6.6}$\cite{koy} suggests that the influence of Pr is weaker
in the underdoped region of the phase diagram than at optimal doping. In
contrast, our value of $\Delta T_{c}/y$ (0.94 K/\%) is more than twice this
previous report (triangles, Fig.~6), and has about the same
magnitude as the $T_{c}$ reduction for YBCO$_{7}$:Pr. To investigate the
role of Pr substitution in the deoxygenated samples, we studied the $^{17}$O
NMR in two Pr$_{y}$O$_{6.6}$ samples corresponding to the highest Pr
concentrations studied at full oxygenation in the previous section, $y=0.1$
and $0.2$.

In the pure samples, the variation of the NMR shifts with oxygen content is
much weaker in the underdoped regime than near optimal doping (see Fig.~2).
Only a small reduction of the difference of $^{17}$K$_{c}$
between 100 and 300 K is observed, from $x=0.65$ to $x=0.45$
so that one cannot obtain a very accurate determination of the counterdoping
from NMR.\ The data for both the $y=0.1$ and $0.2$ samples appear to be
simply shifted versions of that for the unsubstituted $x=0.6$ sample (Fig.~8).
This observed change in the $T$ independent shift contribution appears too large
to be of quadrupolar origin (see above), especially since no such effect was
found in the O$_{7}$ composition. It is also large relative to the ($T$
independent) orbital shift, about 100 ppm in YBCO$_{7}$. Such behaviour
precludes a quantitative estimate of the underdoping effect since the
calibration of Section III.B does not apply. However, the absence of a
reduction of the difference of $^{17}$K$_{c}$ between 100 K and 300 K with
increasing Pr content suggests that again counterdoping is insufficient to
explain the $\sim $20K decrease of $T_{c}$ observed for $y=0.2$.

\begin{figure}[t]
\centerline{\includegraphics[width=7cm]{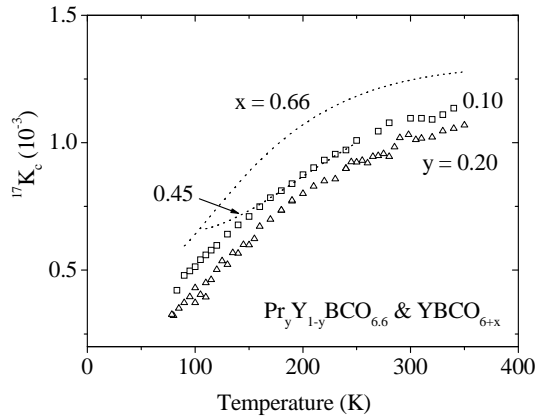}}
\vskip 0.2cm
\caption{$^{17}$K$_{c}$ for various Pr concentrations $y$ in deoxidized Pr$%
_{y}$Y$_{1-y}$Ba$_{2}$Cu$_{3}$O$_{6.6}$ (points) together with curves for
deoxidized YBa$_{2}$Cu$_{3}$O$_{6+x}$ (curves). }
\label{o66shif}
\end{figure}

Let us now consider the Pr-induced broadening of the $^{17}$O line, which
has been linked for YBCO$_{7}$ with magnetic effects associated with the
Pr.\ This extra linewidth is quite small for the sample with $y=0.1$.\ For $%
y=0.2$ it is larger and increases very rapidly at low $T$ in the underdoped
regime (Fig.~9), much faster than a Curie law. This is
reminiscent of the situation encountered in underdoped YBCO$_{6.6}\,$for
substitution of the plane Cu by a magnetic ion, such as Ni \cite{plimp},
where the rapid $T$ variation of the width has been shown to be a
consequence of the $T $ dependent enhanced magnetic response $\chi ^{\prime
}\left( {\bf q}\right)$ of the underdoped planes. There is no reason to
expect this characteristic of the underdoped planes to be altered by Pr
substitution, so a detailed comparison with the case of Ni may be
instructive. For Ni/plane Cu substitution, the broadening $\Delta \nu $%
(100K) of the plane oxygen NMR was found 3 to 4 times larger than near
optimal doping, while the magnitute of the susceptibility of Ni does not
change significantly with oxygen content. However, here $\Delta \nu $(100K)
for the same concentration $y$ of Pr is {\it smaller} for Pr$_{y}$:O$_{6.6}$
compared to Pr$_{y}$:O$_{7}$. As for the Pr effective moment, its magnitude
is not significantly modified from $x=1$ to $x=0.6$ by the changes in
crystalline electric field at the Pr site\cite{jaya,hilsch,uma96}. Therefore
the large decrease in linewidth compared to the Pr$_{y}$:O$_{7}$ linewidth
data indicate that the coupling of the Pr moment to the planes is reduced in
the underdoped state.\ We discuss this point in the next section together
with the data in the optimally doped composition.

\begin{figure}[t]
\centerline{\includegraphics[width=7cm]{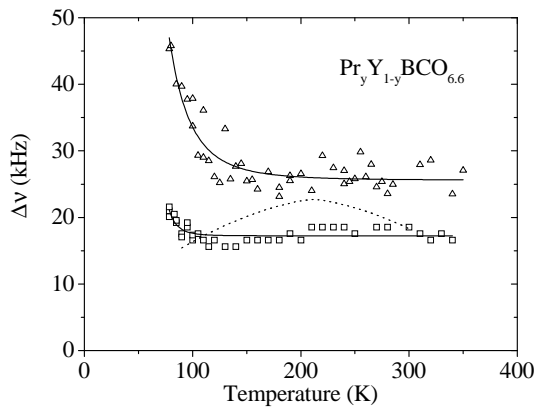}}
\vskip 0.2cm
\caption{ The widths of the O(2,3) mainline, and the temperature dependence
of the width in pure YBa$_{2}$Cu$_{3}$O$_{6.6}$ (dashed). $y=$0.1(squares),
0.2(triangles). }
\label{o66wid}
\end{figure}

\section{Discussion}

The data presented here demonstrate unambiguously that in
Pr$_{y}$:O$_{7}$ Pr acts not only as a counterdopant (reducing $n$),
but also as a magnetic defect. This is reflected in the long-range oscillatory spin
polarisation which gives rise to the broadening of the plane oxygen NMR.
The magnetic defect couples strongly to the CuO$_{2}$ planes in YBCO$_{7}$,
while the coupling apparently decreases for deoxygenated samples.

The effective local magnetic moment of the Pr ion couples directly to
the planes, in contrast to the moments of other rare earth ions substituted
on the Y site. Indeed the ordering temperature of the Pr moments (17 K \cite
{keb89}) in pure nonmetallic PrBaCuO$_{7}$ is much larger than
the ordering temperature of Gd in GdBaCuO$_{7}$, which is a high Tc
superconductor\cite{dun87}, indicating stronger superexchange paths through
the planes for Pr than for Gd. Such a coupling via, for example,
hybridization of Pr with its near neighbor oxygens as suggested by
Fehrenbacher and Rice\cite{FR} would also result in a magnetic pair breaking
effect.

Additionally such a hybridization should change the superexchange interaction between
the Cu hole spins adjacent to the Pr ion. Such a local perturbation in the Cu-Cu
interactions may induce an extended magnetic defect in the plane, as does
any in-plane perturbation. This magnetic defect would contribute both to the
induced spin polarization and the $T_c$ depression, similar to Zn and Ni impurities\cite{indmom}.

We do not at this stage have enough detailed information regarding the local
magnetic properties around the Pr impurities to distinguish between these two
contributions, as the NMR line of Fig.~1 is comprised only
of distant oxygen nuclei.

We can, however, qualitatively consider the likely evolution of these
effects with decreasing $x$, that is deoxygenation. The
established increase of the Pr-O(2,3) distance due to removal of chain oxygen
\cite{collin,uma96} is compatible with a weaker effect of Pr in the underdoped
regime. In particular it could cause {\em both} a decrease of the exchange
coupling of the Pr moment to the planes {\em and} a weaker induced magnetic
defect in the planes. The former may be responsible for the monotonic
decrease of the Pr ordering temperature to about 11 K as $x\rightarrow 6$
found in PrBCO$_{x}$\cite{keb89,uma96,mal94}.

Another important method of characterizing the Pr defect is through
its effect on charge transport, i.e. the residual resistivity $\rho_{0}$.
The anomalous behaviour of in-plane Zn defects\cite{fuku}
and radiation induced disorder\cite{flor} have been studied in detail,
and the correlation between the reduction in T$_c$ and the increase in
$\rho _{0}$ has been found to depend strongly on $n$.
Without resorting to a microscopic formulation of the impurity scattering or
its effect on T$_c$,
it is of interest to ask whether the Pr defect behaves qualitatively differently to
the in-plane defects\cite{dwav}.
Until now such a comparison was not possible as the counterdoping effect could
not be separated from the scattering.
Our measurements provide $\Delta $T$_{c}=\Delta $T$_{c}^{\text{ex}}$ and $n$
at full oxygenation, but transport data on high quality single crystals is lacking.
Some measurements on Pr$_{y}$:O$_{7}$ have been reported and compared to radiation
damage\cite{sun}.
However, we find that $\rho _{0}$ from Ref\cite{sun} is unreasonably large.
Indeed for a 10\% reduction of T$_{c}$, $\rho _{0}\approx $ 5600 $\Omega /$
plane while for Zn it corresponds to $\Delta \rho _{0}<1000$ $\Omega /$plane.
In order to see if this difference indicates that Pr is quite distinct from Zn,
it will be necessary to confirm the reported values of $\rho_0$.
Other techniques, such as microwave and optical conductivity,
that yield the impurity contribution to the carrier scattering rate are
also of interest in a better quantitaive analysis of
$\Delta $T$_{c}^{\text{ex}}$.

\section{Conclusion}

In conclusion, from measurements of planar $^{17}$O NMR, we are able to
deduce the net hole doping in the planes accurately in Pr$_{y}$:O$_{7}$.\
Counterdoping by Pr is observed at Pr$_{y}$:O$_{7}$, but is not sufficient
to explain the reduction of T$_{c}$. The rate of counterdoping (summarized
by $x^{\text{eq}}\approx 1-0.86y$) is important both as a constraint on
theories of the T$_{c}$ suppression and as a means of interpreting the
results of other experiments in which the doping level is not directly
measured. We suggest that the excess loss of T$_{c}$ is related to
scattering of the remaining mobile holes in the CuO$_{2}$ planes by the Pr defects.
This might be confirmed by obtaining accurate estimates of the
residual resistivity due to Pr and reliably obtaining the correlation $%
\Delta $T$_{c}^{\text{ex}}(n,\Delta \rho _{0})$. It has been demonstrated
that the Pr defect induces an in-plane magnetic response which is due
to the coupling of the Pr moment itself to the planes and to the local
modification of superexchange between copper holes near Pr. In Pr$_{y}$:O$%
_{6.6}$, the smaller influence of the magnetic perturbation can be
attributed to the increase of the Pr-O distance .

\section{Acknowledgements}

WAM gratefully acknowledges the support of CIES, France and NSERC, Canada.
We thank S. Uchida, M. B. Maple, and K. Koyama for various communications.



\begin{references}
\bibitem{widnot}  For $y \geq 0.04$, the Pr concentration is far from the
dilute limit (e.g. at $y=0.1$, the mean distance between impurities is on
the order of 2 lattice spacings), and the nonlinear $y$ variation of the
Curie constant evident in the figure may be a manifestation of this. The
dilute-limit expression {\it overestimates} the dipolar width for higher
concentrations, so the conclusion that dipolar fields are too weak to
explain the broadening is unaffected.
\end{references}
\end{document}